\documentclass{aa}
\usepackage{graphicx}
\usepackage{txfonts}
\usepackage{natbib}
\begin{document}
\voffset=-1.0cm
\newcommand{\be}{\begin{equation}}
\newcommand{\ee}{\end{equation}}
\newcommand{\bea}{\begin{eqnarray}}  
\newcommand{\eea}{\end{eqnarray}}

\title{Models of magnetized neutron star atmospheres: thin atmospheres
and partially ionized hydrogen atmospheres \\ with vacuum polarization}

\author{
V.~Suleimanov\inst{1,2},
A.~Y.~Potekhin\inst{3,4},
K.~Werner\inst{1}}

\offprints{V.~Suleimanov}
\mail{e-mail: suleimanov@astro.uni-tuebingen.de}

\institute{
Institut f\"ur Astronomie und Astrophysik, Kepler Center for Astro and
Particle Physics, Universit\"at T\"ubingen, Sand 1, 72076 T\"ubingen, Germany
\and
Kazan State University, Kremlevskaja str., 18, Kazan 420008, Russia
\and
Ioffe Physical-Technical Institute,  Polytekhnicheskaya str., 26, St.
Petersburg 194021, Russia
\and
Isaac Newton Institute of Chile, St.~Petersburg Branch, Russia}

\date{Received XX Xxxxx 2009 / Accepted XX Xxxx XXXX}

   \authorrunning{Suleimanov, Potekhin \& Werner}
   \titlerunning{Models of magnetized neutron star atmospheres}

\abstract
{Observed X-ray spectra of some isolated magnetized neutron stars 
display absorption features,
sometimes interpreted as ion cyclotron lines. Modeling the observed spectra is
necessary to check this hypothesis and to evaluate neutron star parameters.} 
{We develop a computer code for modeling magnetized neutron star atmospheres
in a wide range of magnetic fields ($10^{12} - 10^{15}$ G) and
effective temperatures ($3 \times 10^5 - 10^7$ K).
Using this code, we study the possibilities to explain the soft X-ray
spectra of isolated neutron stars
by different atmosphere models.}
{The atmosphere is assumed to consist either of
fully ionized electron-ion plasmas or
of partially ionized hydrogen.
Vacuum resonance and partial mode conversion are taken into account.
Any inclination of the magnetic field relative
to the stellar surface is allowed.
We use modern opacities of fully or partially ionized plasmas
in strong magnetic fields and solve the coupled radiative transfer equations
for the normal electromagnetic modes in the plasma.}
{Spectra of outgoing radiation are calculated for various atmosphere models:
fully ionized semi-infinite atmosphere, thin atmosphere, 
partially ionized hydrogen atmosphere,
or novel ``sandwich'' atmosphere
(thin atmosphere with a hydrogen layer above a helium
layer).
Possibilities of applications of these results are discussed.
 In particular, the outgoing spectrum
using the ``sandwich'' model is constructed.
Thin partially ionized hydrogen atmospheres with
vacuum polarization are shown
to be able to improve
 the fit  to the observed spectrum
of the nearby isolated neutron star RBS 1223 (RX J1308.8+2127).}
{}

\keywords{radiative transfer -- methods: numerical --
 stars: neutron -- stars: atmospheres -- X-rays: stars --
 stars: individual: RX J1308.8+2127}

\maketitle
%

\section{Introduction}

In the last two decades, several new classes of neutron stars (NSs) have
been discovered by X-ray observatories. They include X-ray dim isolated NSs
(XDINSs), or Magnificent Seven (see review by \citealt{Haberl:07}), 
central compact objects (CCOs) in supernova 
remnants \citep{Pavlovetal:02a,Pavlovetal:04}, anomalous X-ray pulsars 
and soft-gamma repeaters (AXPs and SGRs; see reviews by 
\citealt{Kaspi:07,Mereghettietal:07,Mereghetti:08}). 
The NSs in the 
last two classes have superstrong magnetic fields ($B \ga 10^{14}$ G) and 
are commonly named magnetars. The XDINSs can have 
$B \sim \mbox{a few} \times 10^{13}$~G, 
as evaluated from period changes and from absorption features in the observed 
spectra, if they are interpreted as ion cyclotron lines (see reviews by 
\citealt{Haberl:07,vKK:07}). 
 
These NSs are relatively young with ages $\le 10^6$ yr and 
sufficiently hot ($T_{\rm eff} \sim 10^6 - 10^7$ K) to be observed as soft X-ray 
sources. 
The thermal spectra of 
these objects can be described by blackbody spectra with (color) temperatures 
from 40 to 700 eV (see, for example, \citealt{Mereghettietal:02}). 
Some of the XDINSs and CCOs in supernova 
remnants have one or more absorption features in their X-ray spectrum at the 
energies 0.2 -- 0.8 keV \citep{Haberl:07}. 
 The central energies of these features appear to be 
harmonically spaced \citep{Sanwaletal:02,Swopeetal:07,vKK:07,Haberl:07}. The optical 
counterparts of some XDINSs 
are also known (see review by 
\citealt{Mignanietal:07}), and their optical/ultraviolet fluxes are a few 
times larger than the  
blackbody extrapolation of the X-ray spectra 
\citep{Burwitzetal:01,Burwitzetal:03,Kaplanetal:03,Motchetal:03}. 
 
The XDINs are nearby objects, and parallaxes of some of them can 
be measured \citep{Kaplanetal:02a}. Therefore, they give a good possibility to 
measure the NS 
radii, yielding useful information on the equation of state 
(EOS) for the NS inner core \citep{Trumperetal:04,LP07}, 
which is 
one of the most important problems in the NS physics. For example, the 
EOS is necessary 
for computations of templates of gravitational wave signals which arise 
during neutron stars merging (e.g. \citealt{Baiottietal:08}). 
 
For a sufficiently accurate evaluation of NS radii, 
a good model of the NS surface radiation for 
the observed X-ray spectra fitting is necessary. The isolated NS surface 
layers can either be condensed or have a plasma envelope \citep{Lai.Salpeter:97, 
Lai:01}.  In the latter case, the outer envelope layer forms an NS 
atmosphere. The structure and emergent spectrum of this atmosphere can be 
computed by using stellar model atmosphere methods \citep[e.g.\ 
][]{Mihalas:78}.  Such modeling has been performed by many scientific 
groups for NS model 
atmospheres without magnetic field \citep{Romani:87,Zavlinetal:96, 
Rajagopal.Romani:96, Werner:00, Gansickeetal:02, Ponsetal:02} 
and by several groups 
for models with strong ($B \ga 10^{12}$ G) magnetic fields 
\citep{Shibanovetal:92, Rajagopaletal:97, Ozel:01, Ho.Lai:01, Ho.Lai:03, 
Ho.Lai:04,vAL:06}. 
These model spectra were used to fit the observed isolated 
 NS X-ray spectra (see review by \citealt{Zavlin:09}). 
 
Modeling of magnetized NS star atmospheres is 
based on the theory of electromagnetic wave propagation 
in a magnetized plasma in two normal modes, extraordinary (X) and ordinary (O) ones 
\citep{Ginzburg:70, Mezharos:92}, and on the methods of opacity calculations for these 
two modes \citep{Ventura:79, Kaminkeretal:82, Kaminkeretal:83}.  
Methods of fully 
ionized model atmospheres  modeling are well 
developed (see, e.g., \citealt{Zavlin:09} 
for references). Partially ionized hydrogen 
atmospheres have been modeled 
\citep{Potekhinetal:04, Ho.Lai:04, Hoetal:08}, 
using the opacity and EOS calculations 
by \cite{PCh:03,PCh:04}, 
which accurately take into account the motional Stark effect 
and plasma nonideality effects in quantizing magnetic fields.  
Mid-$Z$ element atmospheres for strongly 
magnetized NSs have been modeled by 
\citet{MH:07}, who treat the 
motional Stark effect using a perturbation 
theory \citep{PM:93} valid at relatively low $T$. 
 
For magnetar atmospheres, 
polarization of the vacuum can be significant, which was 
studied by \citet{PG:84} and recently by \cite{Lai.Ho:02, Lai.Ho:03}. 
Model atmospheres 
with partial mode conversion due to the vacuum 
polarization have been computed 
by \citet{Ho.Lai:03} 
and \citet{vAL:06}. 
 
If the temperature is sufficiently  low or the magnetic field is
sufficiently strong, the thick atmosphere 
can be replaced by a condensed surface 
 \citep{Lai.Salpeter:97,Lai:01,M07} 
as a result of the plasma phase transition (cf.~\citealt{PCS:99,PCh:04}). 
Emission and absorption properties of such surfaces in 
strong magnetic fields have been studied by \citet{TZD:04,vAetal:05,PAMP05}. 
 
In recent years, evidence has appeared that some of the XDINSs 
may have a ``thin'' atmosphere above the condensed surface. 
Such atmosphere could be optically thick to low-energy photons and optically 
thin to high-energy photons. 
\citet{Motchetal:03} fitted the spectrum of RX J0720.4$-$3125 
using a nonmagnetic thin hydrogen atmosphere model, 
and \citet{Hoetal:07} fitted the spectrum of RX J1856.4$-$3754 
using a magnetic, partially ionized, thin hydrogen atmosphere model. 
 
In this paper we present a new computer code for model atmospheres of 
magnetized NSs and some new results, which were obtained by using this 
code. The code is applicable to modeling of fully ionized 
NS atmospheres and partially ionized hydrogen atmospheres  
accounting for the partial 
mode conversion due to the 
vacuum polarization. The
magnetic field can be normal or inclined to NS surface. 
We use this code to study various thin atmospheres above 
a condensed NS surface. 
In particular, we consider a novel ``sandwich'' atmosphere 
model, where a thin atmosphere is composed of a hydrogen 
layer above a helium layer. 
 We also discuss the possibility to apply our thin 
partially ionized hydrogen model atmospheres with vacuum polarization 
to fit the spectrum of RBS 1223.
Our numerical method is outlined in \S\,\ref{s:methods}; 
Results are presented and discussed in \S\,\ref{s:results}. 
Conclusions are summarized in \S\,\ref{s:conclusions}.

\section{Method of atmosphere structure calculations} 
\label{s:methods} 
 
\begin{figure} 
\includegraphics[width=\columnwidth]{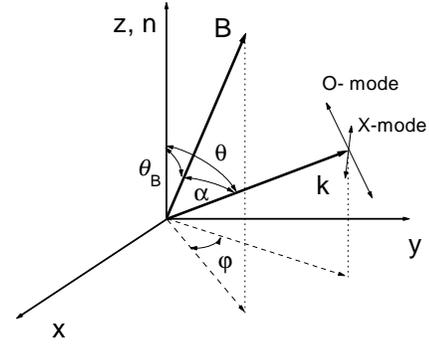} 
\caption{\label{f:fig1} 
Geometry of the radiation transfer. 
 } 
\end{figure}

We compute model 
atmospheres of hot, magnetized NSs subject to the constraints 
of hydrostatic and radiative equilibrium assuming planar geometry 
and homogeneous magnetic field. 
There are two versions of the code. 
In the first one, we consider the magnetic field 
$\vec{B}$ perpendicular to the surface.  In 
this case the angle $\alpha$ between $\vec{B}$ and a radiation 
wave vector $\vec{k}$ is equal to the angle $\theta$ between 
$\vec{k}$ and the normal $\vec{n}$ to the surface 
(see Fig.~\ref{f:fig1}). 
It is the simplest case, 
because opacities depend on $\alpha$, and the geometry of radiation 
propagation depends on $\theta$.  In the second version, the angle 
$\theta_B$ between $\vec{B}$ and $\vec{n}$ is arbitrary, 
and calculations are more expensive. In this case the 
opacities depend not only on the polar angle 
$\theta$, but also on the azimuthal angle 
$\varphi$ between the projections of $\vec{B}$ and $\vec{n}$ 
onto the stellar surface, therefore it is necessary to solve radiation 
transfer equations for a significantly larger number of directions. 
 
The model atmosphere structure for a NS with effective temperature $T_{\rm 
eff}$, surface gravity $g$, magnetic field $B$, and given chemical composition 
is described by the following 
set of equations: 
\begin{enumerate}\setlength{\itemsep}{1ex} 
\item 
The hydrostatic equilibrium 
equation 
\be \label{e:hyd} 
  \frac {\mathrm{d} P_{\rm g}}{\mathrm{d}m} = g - g_{\rm rad}, 
\ee 
where 
\be 
    g=\frac{GM_{\rm NS}}{R^2_{\rm NS}\sqrt{1-R_{\rm S}/R_{\rm NS}}}, 
\ee 
\be 
 g_{\rm rad} =  \frac{1}{c} \sum_{i=1}^2 \int_0^{\infty}\!\!\! \mathrm{d}\nu \, 
\int_0^{2\pi}\!\!\! \mathrm{d}\varphi \, 
\int_{-1}^{+1}\!\!\! (k_{\nu}^i+\sigma_{\nu}^i) \, \mu \, I_{\nu}^i(\mu, \varphi) 
\, \mathrm{d}\mu 
\ee 
allows for the radiation pressure, 
and $R_{\rm S}=2GM_{\rm NS}/c^2$ is the Schwarzschild radius of the NS. 
Here $\mu=\cos \theta$, \,$I_{\nu}^i(\mu, \varphi)$ is the 
specific intensity in mode $i$, $P_{\rm g}$ is  the gas pressure, and the column density $m$ is 
determined as 
\be 
     \mathrm{d}m = -\rho \, \mathrm{d}z \, . 
\ee 
The variable $\rho$ denotes the gas density and $z$ is the vertical distance. 
 Of course, radiation pressure is unimportant in the models
presented below,
but it can be more significant at higher effective temperatures.

\item 
The radiation transfer equations for the two modes 
\be \label{rtr} 
   \mu\frac{\mathrm{d} I_{\nu}^i}{\mathrm{d} \tau_{\nu}^i} 
     =  I_{\nu}^i - S_{\nu}^i 
\ee 
where 
\bea \label{sf} 
& & S_{\nu}^i = \frac{k_{\nu}^i}{k_{\nu}^i+\sigma_{\nu}^i} \frac{B_{\nu}}{2} 
+  \\ \nonumber 
& & 
   \frac{1}{2\pi} \, \frac{1}{k_{\nu}^i+\sigma_{\nu}^i} \sum_{j=1}^2 \, 
\int_0^{2\pi}\!\! \mathrm{d}\varphi' \, \int_{-1}^{+1}\!\! 
\sigma_{\nu}^{ij}(\mu,\varphi;\mu',\varphi')\, I_{\nu}^j (\mu',\varphi')\, 
\mathrm{d}\mu' 
\eea 
is the source function, 
$B_{\nu}$  is the blackbody (Planck) 
intensity, and the optical 
depth $\tau_{\nu}^i$ is defined as 
\be 
    \mathrm{d}\tau_{\nu}^i = (k_{\nu}^i+\sigma_{\nu}^i) \, \mathrm{d}m. 
\ee 
Here, the true absorption and electron scattering opacities $k_{\nu}^i$ and $\sigma_{\nu}^i$ 
depend on $\mu$ and $\varphi$. Specific intensity in given direction and mode 
can be scattered in some other direction and in both modes
\be \label{scats}
     \sigma_{\nu}^i(\mu, \varphi) = \frac{1}{2\pi} \sum_{j=1}^2 \, 
\int_0^{2\pi}\!\! \mathrm{d}\varphi' \, \int_{-1}^{+1}\!\! 
\sigma_{\nu}^{ij}(\mu, \varphi; \mu', \varphi' ) \, \mathrm{d}\mu'
\ee
 
\item 
The energy balance equation 
\bea  \label{econs} 
& & \frac{1}{2\pi} \, \sum_{i=1}^2  \, \int_0^{\infty} \mathrm{d}\nu  \, 
\int_0^{2\pi} \mathrm{d}\varphi  
\nonumber\\ & & 
\times  \int_{-1}^{+1} \left( (k_{\nu}^i 
 +\sigma_{\nu}^i) I_{\nu}^i(\mu,\varphi) - \eta_{\nu}^i(\mu,\varphi) 
\right) \, \mathrm{d}\mu = 0 
\eea 
with emissivity 
\bea 
\eta_{\nu}^i(\mu,\varphi) &=& 
 \frac{1}{2\pi} \, \sum_{j=1}^2 \, 
\int_{0}^{2\pi}\!\!\! \mathrm{d}\varphi' \, \int_{-1}^{+1}\!\!\! 
\sigma_{\nu}^{ij}(\mu,\varphi;\mu',\varphi')\, I_{\nu}^j (\mu',\varphi') \, 
\mathrm{d}\mu' 
 \nonumber \\ 
& + & 
k_{\nu}^i \, \frac{B_{\nu}}{2} \,. 
\label{e:eta} 
\eea 
\end{enumerate} 
 
Equations (\ref{e:hyd}) -- (\ref{e:eta}) 
must be completed by the EOS and the charge 
and particle conservation laws. In the code two different cases of these 
 laws are considered. 
 
\begin{figure} 
\includegraphics[width=1.0\columnwidth]{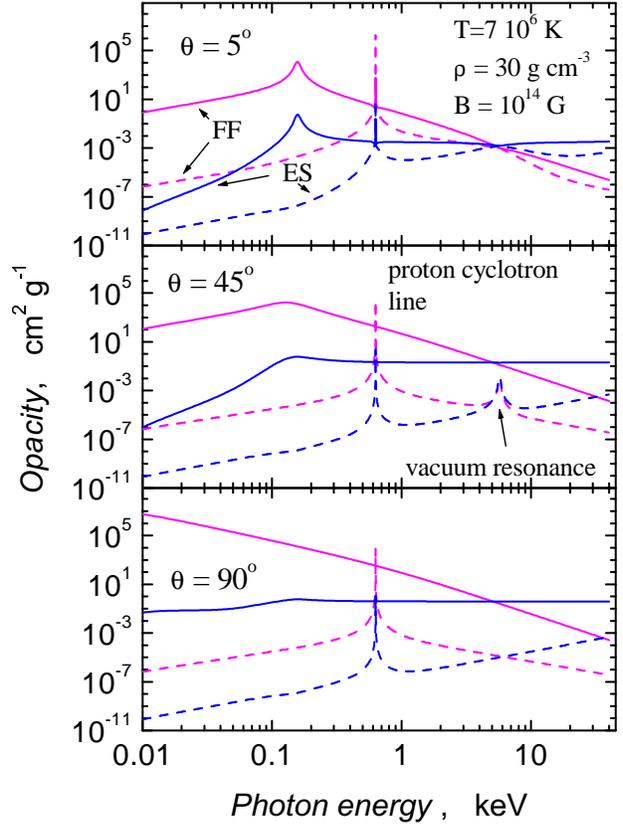} 
\caption{\label{f:fig1a} 
Dependence of the free-free and electron scattering opacities in two 
modes on the photon energy at different angles 
in a fully ionized hydrogen plasma. The proton cyclotron and 
vacuum resonances are also shown. The plasma temperature is 7$\times 10^6$ 
K, the plasma density is 30 g cm$^{-3}$, the magnetic field strength is 
$10^{14}$ G. 
} 
\end{figure}

\begin{figure} 
\includegraphics[width=1.0\columnwidth]{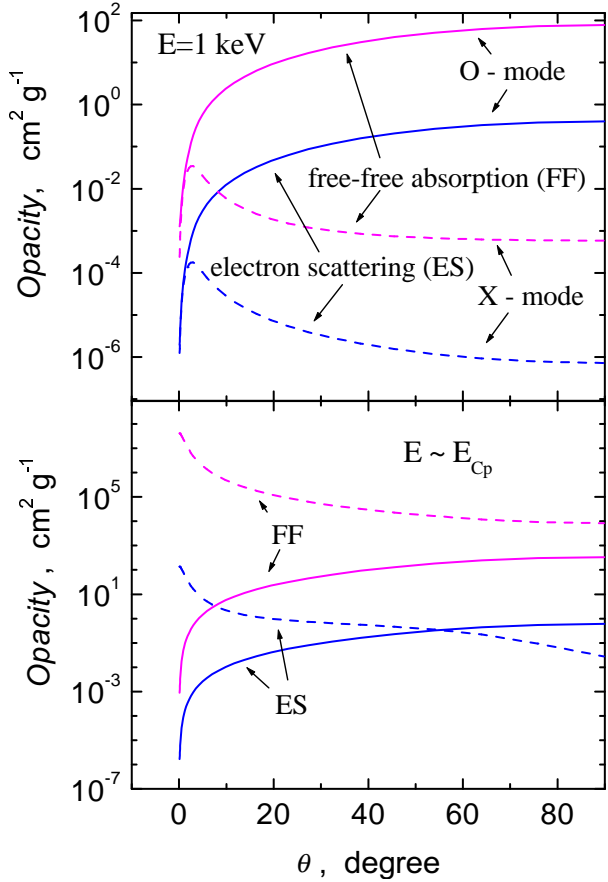} 
\caption{\label{f:fig1b} 
Dependence of the free-free and electron scattering opacities in two 
modes on the angle between the magnetic field lines and the 
direction of photon propagation for two photon energies: 1 keV and at 
the proton cyclotron energy.  Plasma parameters and magnetic field 
strength are the same as in Fig.~\ref{f:fig1a}.} 
\end{figure}

In the first (simplest) case, 
 a fully ionized atmosphere 
are calculated. Therefore, the EOS is 
the ideal gas law 
\be   \label{gstat} 
    P_{\rm g} = n_{\rm tot} kT \,, 
\ee 
where $n_{\rm tot}$ is the number density of all particles. 
Opacities are calculated in the same way as in the paper by 
\cite{vAL:06} (see references therein for the background theory 
and more sophisticated approaches). 
The vacuum polarization effect is taken into account 
following the same work. 
Examples of opacities as functions of photon energy 
and angle $\theta$ 
for a magnetized electron-proton plasma are shown 
in Figs.~\ref{f:fig1a} and \ref{f:fig1b}. 
 
The second considered case is a partially ionized hydrogen atmosphere. 
In this case the EOS and the corresponding 
opacities are taken from tables calculated by 
\citet{PCh:03,PCh:04}. The normal mode polarization vectors 
are taken from the calculations by 
 \cite{Potekhinetal:04}. The vacuum polarization effect 
is also included. 
 
\begin{figure} 
\includegraphics[width=\columnwidth]{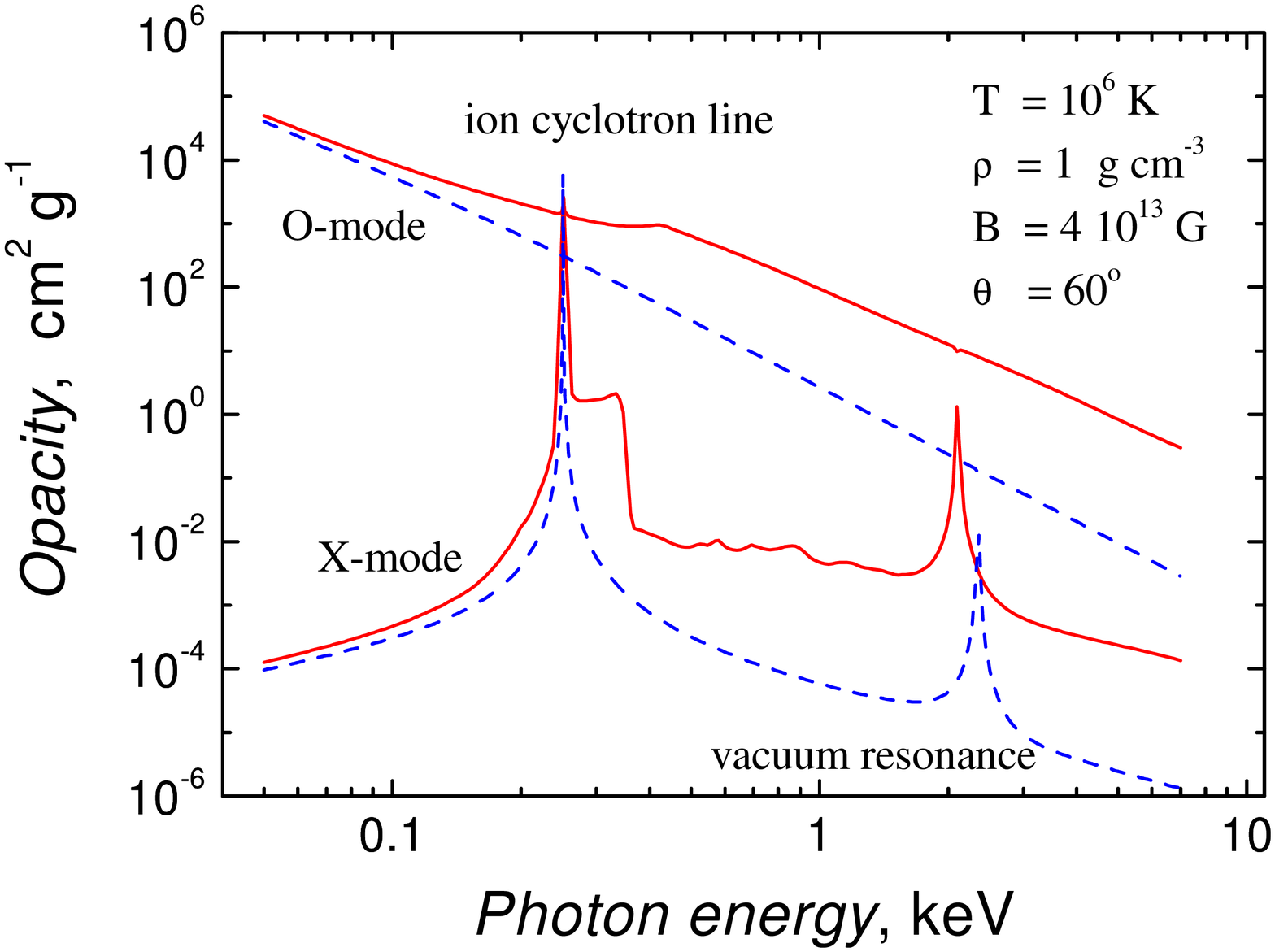} 
\caption{\label{f:fig2} 
Dependence of the true opacities in two 
modes on the photon energy for fully (dashed curves) and partially ionized 
(solid curves) hydrogen. The 
vacuum polarization effect is taken into account. The plasma 
temperature is $10^6$ K, the plasma density is 1 g cm$^{-3}$, the 
magnetic field strength is $4 \times 10^{13}$ G. The angle between  
photon propagation direction and the magnetic field is 60$^{\circ}$.} 
\end{figure} 
 
For solving the above equations and computing the model atmosphere, we 
used a version of the computer code ATLAS \citep{Kurucz:70,Kurucz:93}, 
modified to deal with  strong magnetic fields. A nonmagnetic version of this 
modified code was previously used to model atmospheres of super-soft X-ray sources 
\citep{Swartz.etal:02,Ibragimov.etal:03}, atmospheres of non-magnetized NSs 
\citep{Sul.Wer:07, Rauchetal:08}, and atmospheres of spreading layers 
on the  surface of accreting NSs \citep{Sul.Pout:06}. 
 
The scheme of calculations is as follows.  First of all, the input 
parameters of the model atmosphere  are defined:  $T_{\rm 
eff}$, $g$, $\vec{B}$ and the chemical composition. 
Then a 
starting model using a grey temperature distribution is calculated. 
The calculations are performed with a set of 90 depth points $m_j$ 
 distributed logarithmically in equal steps from $m_1 \approx 
10^{-5}$~g~cm$^{-2}$ to $m_{\rm max} \approx 10^5$~g~cm$^{-2}$ in the case 
of a  semi-infinite atmosphere. It is also possible to calculate 
thin atmospheres with arbitrary values of $m_{\rm max}$. In 
this case the temperature at the inner boundary is considered as 
the temperature of a condensed NS surface. 
 
In the starting model, all number densities and opacities at all 
depth points and all photon energies are calculated. We use 200 
logarithmically equidistant energy points in our computations in the 
range 0.001 -20 keV with 9 additional points near each ion 
cyclotron resonance 
\be \label{ioncyc} 
    E_{\rm ci} = 0.635 \, {\rm keV} \, \frac{B}{10^{14}~{\rm G}} 
\, \frac{Z}{A}, \ee 
where $Z$ is the ionic charge and $A$ the atomic weight in the atomic units. 
If the vacuum resonance is taken into 
consideration, then 
another photon energy grid is used, which is constructed using the ``equal 
grid'' method \citep{Ho.Lai:03}.  In this method every point in the depth 
grid $m_j$ corresponds to the point in the energy grid 
defined by the equation 
\be 
E_i = E_{\rm V} (\rho(m_j)) 
\ee 
(if this energy 
point is in the considered energy range 0.001 -20 keV). 
Here $E_{\rm V}(\rho)$ is the energy of the vacuum resonance at 
given $B$ and $\rho$ (see \citealt{vAL:06}). 
The opacity averaged in the energy interval 
($E_{i-1}$, $E_{i+1}$) is used to avoid opacity overestimation 
in frequency integrals like (\ref{econs}). 
This energy grid is recalculated after every iteration. 
Note that the 
vacuum resonance energy $E_{\rm V}(\rho)$ in a partially ionized plasma 
is shifted relative to its value in a fully 
ionized plasma (see Fig.~\ref{f:fig2}).

The radiation transfer 
equation (\ref{rtr}) is  solved on a set of 40 polar angles $\theta$ 
and 6 azimuthal angles $\varphi$ (in the case of inclined magnetic field) 
by the short characteristic method \citep{Ols.Kun:87}. 
 
We use the conventional condition (no external radiation) at the outer boundary 
\be 
    I_{\nu}^i(\mu < 0, m = m_1) = 0. 
\ee 
The diffusion approximation is used as the inner boundary condition 
\be  \label{in_b_c} 
 I_{\nu}^i(\mu > 0, m = m_{\rm max}) = \frac{B_{\nu}}{2}. 
\ee 
 
The code allows one to take into account the partial mode conversion 
according to \cite{vAL:06}. At the 
vacuum resonance, the intensity in the extraordinary mode 
partially converts 
with the probability $1-P_{\rm jump}$ to the intensity in the ordinary one 
and vice versa 
\be \label{p_conv} 
I_{\nu}^{1,2} \to P_{\rm jump} \, I_{\nu}^{1,2} + (1-P_{\rm jump}) \, 
I_{\nu}^{2,1}, \ee 
where 
\be \label{pjump} 
      P_{\rm jump} = \exp \left[-\frac{\pi}{2}(E/E_{\rm ad})^3\right]. 
\ee 
The value $E_{\rm ad}$ is a function of $E,B,\rho,T,\mu$ (see 
\citealt{vAL:06} for details). 
 
The solution of the radiative transfer equation (\ref{rtr}) is checked 
for the energy balance equation (\ref{econs}), together with the surface 
flux condition 
\be 
    4 \pi \int_0^{\infty} H_{\nu} (m=0) \,\mathrm{d}\nu 
     = \sigma T_{\rm eff}^4 = 4 \pi 
   H_0, 
\ee 
where the Eddington flux at any given depth $m$ is defined as 
\be 
    H_{\nu}(m) = \frac{1}{4\pi} \sum_{i=1}^2 \, \int_{0}^{2\pi} d\varphi \, 
\int_{-1}^{+1} \mu I_{\nu}^i(\mu,\varphi,m) \, \mathrm{d}\mu. 
\ee 
The relative flux error 
\be 
     \varepsilon_{H}(m) = 1 - \frac{H_0}{\int_0^{\infty} H_{\nu} (m) d\nu}, 
\ee 
and the energy balance error as  functions of depth 
\bea 
 & & \varepsilon_{\Lambda}(m) =    \sum_{i=1}^2 \, \int_0^{\infty} 
\mathrm{d}\nu \, \int_0^{2\pi} \mathrm{d}\varphi 
 \nonumber\\ 
& & \times 
   \int_{-1}^{+1} \left( (k_{\nu}^i 
+\sigma_{\nu}^i)  I_{\nu}^i(\mu,\varphi) - \eta_{\nu}^i(\mu,\varphi) 
\right) \, \mathrm{d}\mu 
  \label{econs1} 
\eea 
are calculated. 
 
\begin{figure} 
\includegraphics[width=1.0\columnwidth]{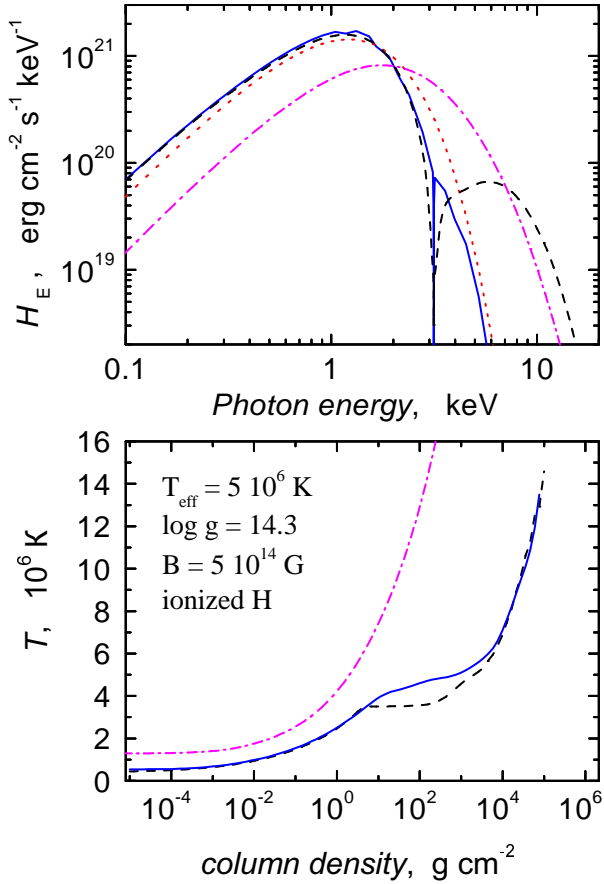} 
\caption{\label{f:fig3} 
Emergent spectra and temperature structures of fully ionized hydrogen 
model atmospheres of neutron stars with $T_{\rm eff} = 5 \times 10^6$ K 
and $\log g = 14.3$.  The models with and without (dash-dotted curves) strong 
 ($B=5 \times 10^{14}$ G) magnetic field as well as with (solid curves) and without (dashed curves) 
vacuum polarization 
 effect are shown.  The blackbody spectrum with the temperature equal 
the effective temperature is also shown in the upper panel (dotted curve). 
} 
\end{figure} 
 
Temperature corrections are then evaluated using three different 
procedures.  The first is the integral $\Lambda$-iteration method, 
modified for the two-mode radiation transfer, based on the energy balance equation 
(\ref{econs}). In this method the temperature correction for a particular 
depth is found from 
\be 
     \Delta T_{\Lambda} = \frac{-\varepsilon_{\Lambda}(m)}{\int_0^{\infty} 
 \Phi_{\nu} \,\mathrm{d}\nu} \,, 
\ee 
where 
\be 
\Phi_{\nu} = \sum_{i=1}^2 \left[ 
(\Lambda_{\nu\mathrm{diag}}^i-1)/(1-\alpha_{\nu}^i\Lambda_{\nu\mathrm{diag}}^i) \right] 
\overline{k}_{\nu}^i (\mathrm{d}B_{\nu}/\mathrm{d}T) 
\ee 
and $\alpha_{\nu}^i=\overline{\sigma}_{\nu}^i/(\overline{k}_{\nu}^i+\overline{\sigma}_{\mu}^i)$. 
Averaged opacities are defined as 
\bea 
    \overline{k}_{\nu}^i = \frac{1}{4\pi} \int_0^{2\pi} 
d\varphi \, \int_{-1}^{+1} k_{\nu}^i(\mu,\varphi) \, 
\mathrm{d}\mu, 
 \\ 
  \overline{\sigma}_{\nu}^i = \frac{1}{4\pi} \int_0^{2\pi} 
\mathrm{d}\varphi 
\, \int_{-1}^{+1} \sigma_{\nu}^i(\mu,\varphi) \, \mathrm{d}\mu. 
\eea 
$\Lambda_{\nu\mathrm{diag}}^i$  is the diagonal matrix element of the $\Lambda$ 
operator depending on the mean optical depth in given mode $i$. 
  The mean optical depth is defined by 
$\mathrm{d}\, \overline{\tau}_{\nu}^i = 
 (\overline{k}_{\nu}^i+\overline{\sigma}_{\mu}^i)\, \mathrm{d}m$ 
(see \citealt{Kurucz:70} for details of 
$\Lambda_{\nu\mathrm{diag}}$ -- $\tau_{\nu}$ dependence). 
This procedure is used in the upper atmospheric layers.  The 
second procedure is the Avrett-Krook flux correction, which uses the 
relative flux error $\varepsilon_{H}(m)$, and it is performed in the deep layers. 
And the third 
one is the surface correction, which is based on the emergent flux 
error (see \citealt{Kurucz:70} for a detailed description). 
 
The iteration procedure is repeated until the relative flux error is 
smaller than 1\% and the relative flux derivative error is smaller 
than 0.01\%. As a result of these calculations, we obtain a 
self-consistent isolated NS  model atmosphere, together with the emergent 
spectrum of radiation. 
 
\begin{figure} 
\includegraphics[width=1.0\columnwidth]{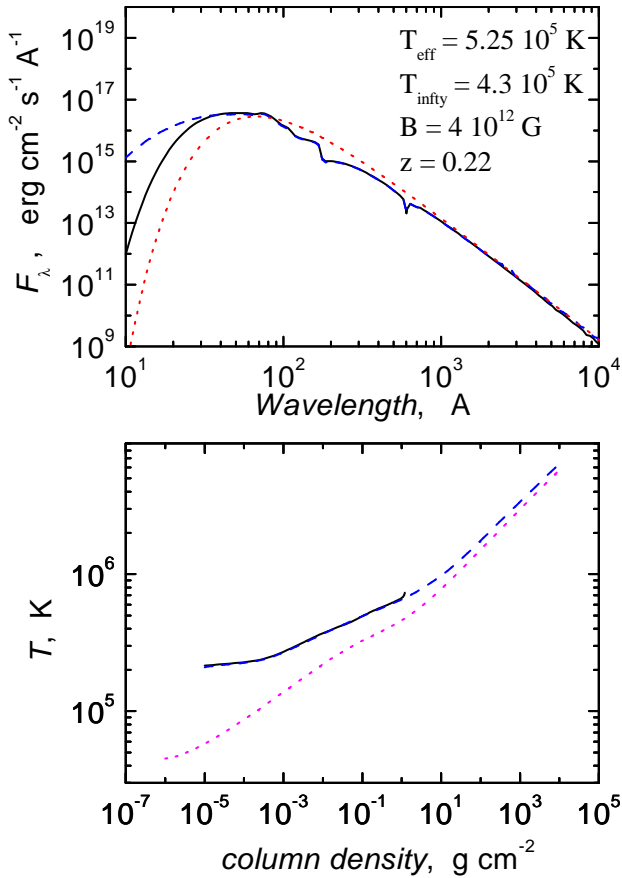} 
\caption{\label{f:fig6} 
{\it Top panel:} Emergent spectra of the 
partially ionized hydrogen semi-infinite (dashed curve) and thin ($\Sigma 
= 1.2 g cm^{-2}$, solid curve) model atmospheres  ($T_{\rm 
eff} = 5.25 \times 10^6$ K, $B=4 \times 10^{12}$ G) together with the 
corresponding blackbody spectrum (dotted curve).  The spectra are calculated with 
gravitational redshift ($z$=0.22) taking into consideration. 
{\it Bottom panel:} Temperature structures of the models from the top panel 
together with the temperature structure of a fully ionized hydrogen model (dotted curve). 
} 
\end{figure}

Our method of calculation has been tested by a comparison to models for magnetized NS 
atmospheres \citep{Shibanovetal:92, Pavlovetal:94, Ho.Lai:01, Ozel:01, 
Ho.Lai:03}.  Model atmospheres with partially ionized 
hydrogen are compared to models computed by \cite{Hoetal:07}.
We have found that our models are 
in a good agreement with these  calculations. Our results are 
presented in 
Fig.~\ref{f:fig3}, where the temperature structures and the emergent spectra 
of  models with and without vacuum polarization are compared to a 
model without magnetic field. In 
Fig.~\ref{f:fig6} we present 
 emergent spectra and temperature structures of 
the semi-infinite and thin model atmospheres with the same parameters 
as used by \cite{Hoetal:07}. 
 
\section {Results} 
\label{s:results} 
 
In this work we use the developed code 
mainly for studies of thin 
atmospheres above a condensed NS surface. In all calculations below we use 
the same surface gravity, $\log g$ = 14.3.

\begin{figure} 
\includegraphics[width=1.0\columnwidth]{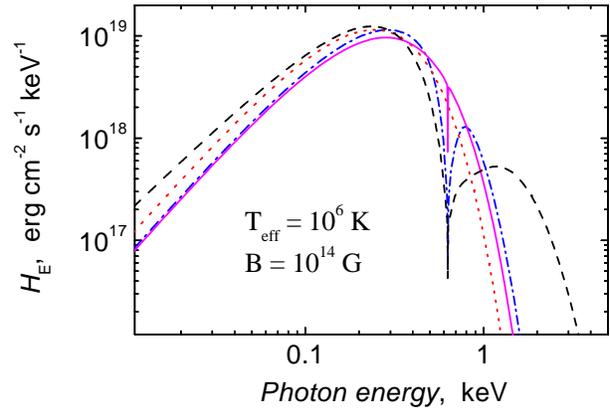} 
\caption{\label{f:fig4} 
 Emergent spectra of the thin fully ionized hydrogen atmospheres above a 
 solid surface with $T_{\rm eff} = 10^6$ K and $B = 10^{14}$ G in 
 comparison to the semi-infinite atmosphere (dashed curve). The 
 spectra of atmospheres with  surface densities $\Sigma$=1 (solid 
curve) and 100 g cm$^{-2}$ (dash-dotted curve) together with the corresponding 
blackbody (dotted curve) are shown. Vacuum polarization effect is not included.
} 
\end{figure}

One of the problem related to magnetars (in particular, AXPs) 
is the lack of any absorption feature at the 
proton cyclotron energy, 
although early models of the magnetized NS 
atmospheres predicted a strong feature 
at the magnetar field strengths 
\citep{Ho.Lai:01,Zaneetal:01}. \citet{Ho.Lai:03} 
suggested that a possible 
solution of this problem is the 
suppression of the cyclotron absorption feature due to the 
vacuum polarization. We confirm this result (see Fig.~\ref{f:fig3}). 
In addition, we 
demonstrate that this 
absorption line is further reduced in a thin atmosphere
 without vacuum polarization. 
Figure~\ref{f:fig4} demonstrates emergent spectra of 
 thin atmospheres without allowance for 
the vacuum polarization effect. 
In the semi-infinite atmosphere, 
a wide proton cyclotron line 
forms in agreement with the results of \citet{Ho.Lai:01,Zaneetal:01}. 
However, the absorption 
feature disappears with decreasing 
the atmosphere surface density $\Sigma$. 
The thin atmosphere is transparent to the 
continuum and absorption line wings, therefore 
the emergent spectrum approaches the 
spectrum of the condensed surface everywhere except for 
a narrow energy band at the 
center of the cyclotron line. 
The width of this absorption 
depends on the atmosphere thickness. 
 
\begin{figure} 
\includegraphics[width=1.0\columnwidth]{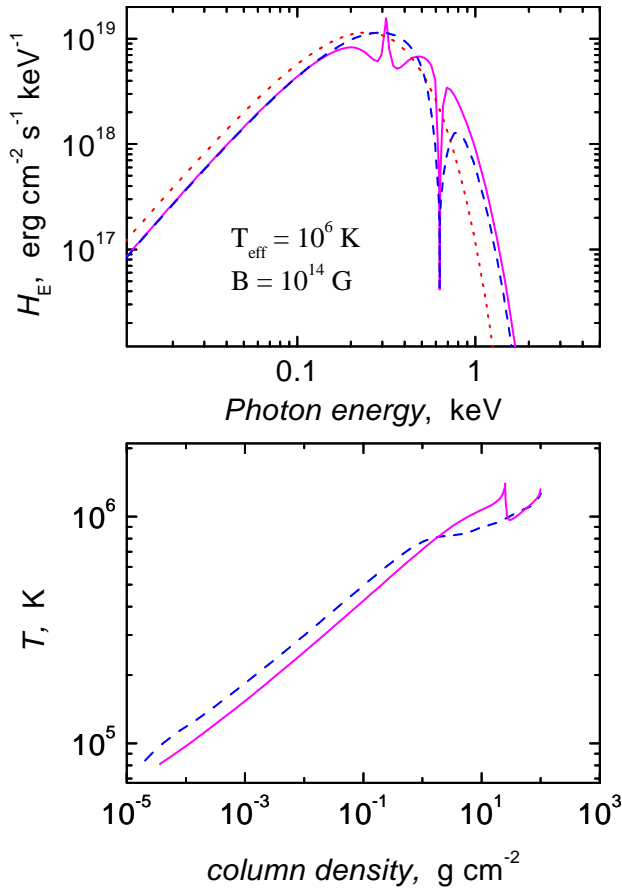} 
\caption{\label{f:fig4a} 
 Emergent spectrum (top panel)  and temperature structure (bottom panel)
 of the ``sandwich'' model atmosphere above a 
 solid surface with $T_{\rm eff} = 10^6$ K and $B = 10^{14}$ G (solid 
curve) in comparison with the thin fully ionized hydrogen atmosphere 
 (dashed curve) with the same parameters.  The surface densities of both 
 model atmospheres are 100 g cm$^{-2}$, in the ``sandwich'' model the H slab 
has 25 g cm$^{-2}$ surface density and the He slab has  75 g cm$^{-2}$. The 
 corresponding blackbody spectrum (dotted curve) is also shown in the top
panel. 
} 
\end{figure} 

Some XDINSs and CCOs show one or two 
 absorption features 
\citep{Haberl:07, Sanwaletal:02, Swopeetal:07}. 
Various hypothesis (so far inconclusive) were 
considered for an explanation  
(see \citealt{MH:07} and references therein). 
Here we 
suggest another one, which we name ``sandwich atmosphere''. 
A thin, chemically layered atmosphere above a condensed NS surface can arise from 
accretion of interstellar gas with cosmic chemical composition. In 
this case, hydrogen and helium quickly 
separate due to the high  
gravity 
(according to \citealt{BBC:02}, the He/H stratification timescale 
can be estimated as 
$\sim \rho_1^{1.3}\,T_6^{-0.3}\,g_{14}^{-2}$~s, 
where $\rho_1=\rho/10$ g~cm$^{-3}$, $T_6=T/10^6$~K, and 
$g_{14}=g/10^{14}$ cm~s$^{-2}$). 
In this ``sandwich atmosphere'' a layer 
of hydrogen is located above a helium slab, 
and the emergent spectrum has two 
absorption features, corresponding to proton and $\alpha$-particle 
cyclotron energies. In Fig.~\ref{f:fig4a} the emergent spectrum for one of 
such models is shown.  The emission feature at the helium absorption line 
arises due to a local temperature bump at the boundary between the helium and 
hydrogen layers. This bump arises due to sharp changing of the plasma density and 
the opacity between helium and hydrogen layers. Clearly, some transition zone with
mixed H/He chemical composition must exist between layers, and this rapid temperature 
change can be reduced. We plan to calculate models with this kind of transition 
zone in future work.

\begin{figure} 
\includegraphics[width=1.0\columnwidth]{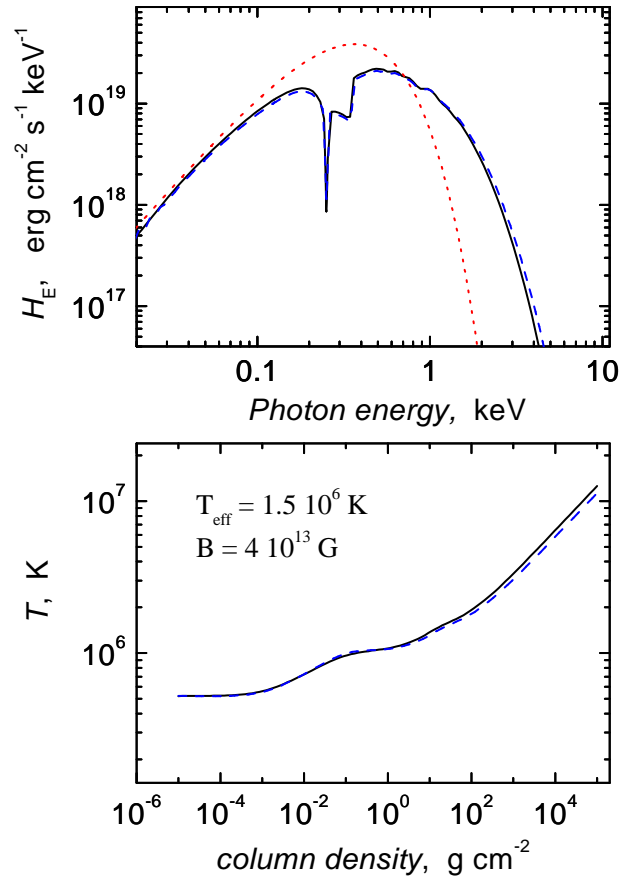} 
\caption{\label{f:fig7} 
Emergent spectra and temperature structures of partially ionized 
 hydrogen model atmospheres  with $T_{\rm eff} = 1.5 \times 
10^6$ K with  inclinations of the magnetic field ($B=4 \times 
10^{13}$ G) to the surface normal $\theta_B$ equal 0$^{\circ}$ (solid 
curves) and 90$^{\circ}$ (dashed curves).  The vacuum polarization effect is not 
included.  The corresponding blackbody spectrum  is also shown in 
the upper panel (dotted curve).  
} 
\end{figure} 
 
For comparison to observations, it is necessary to integrate the local 
model spectra over the NS surface. The effective 
temperature and magnetic field strengths are not uniform 
over the NS surface, and generally the magnetic field is 
not perpendicular to the surface (see \citealt{Hoetal:08}). Therefore, 
it is necessary to compute model atmospheres with inclined magnetic 
field. This possibility is included in our code. For example, 
Fig.~\ref{f:fig7} shows 
spectra and temperature structures of model atmospheres 
with magnetic field  perpendicular and  parallel to the NS surface. 
 
\begin{figure} 
\includegraphics[width=1.0\columnwidth]{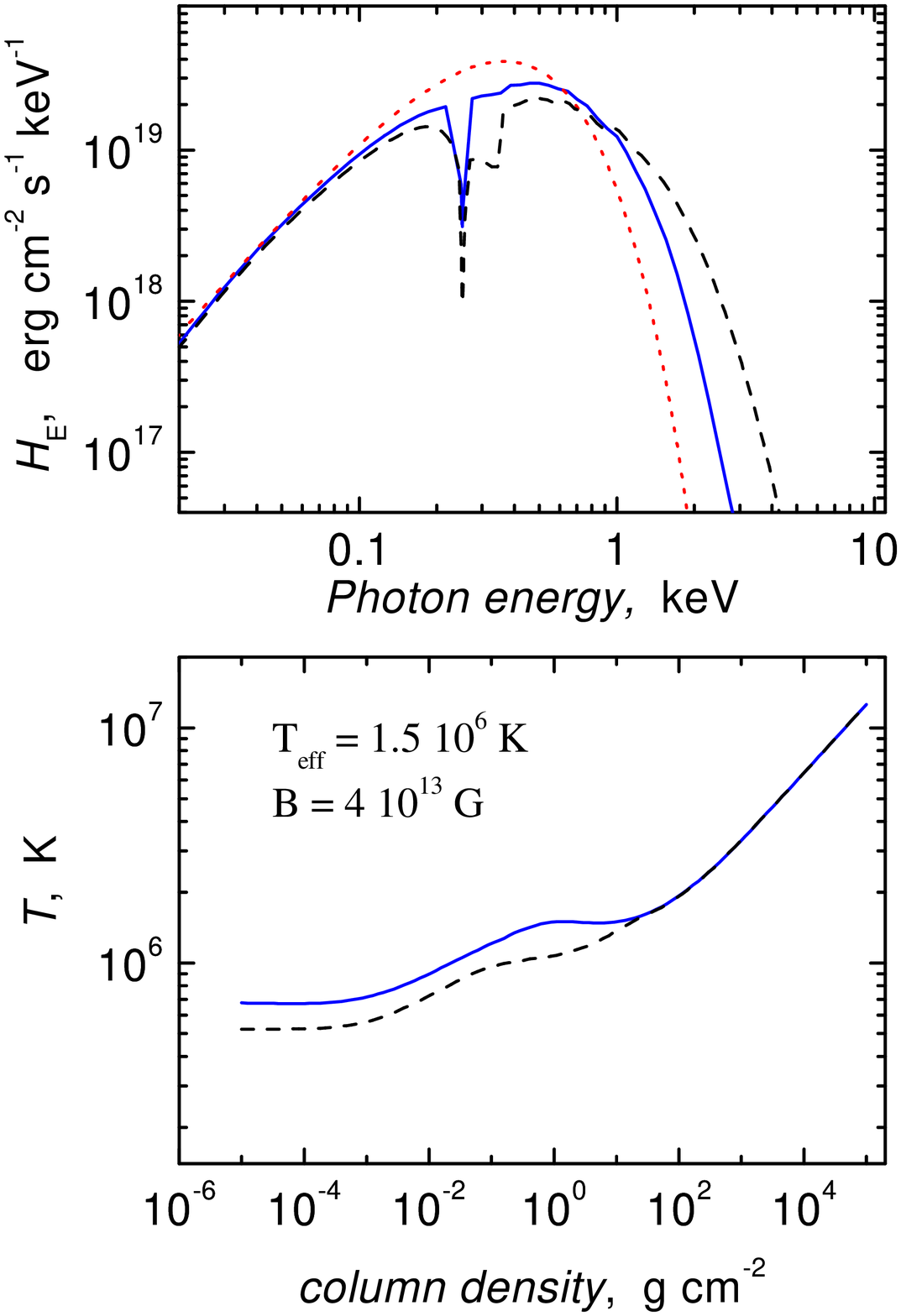} 
\caption{\label{f:fig8} 
Emergent spectra and temperature structures of the partially ionized 
 hydrogen model atmospheres  with $T_{\rm eff} = 1.5 \times 
10^6$ K with (solid curves) and without (dashed curves) vacuum polarization 
effect (with the partial mode conversion). The magnetic field strength is $B=4 \times 
10^{13}$ G. The corresponding blackbody spectrum is also shown in the 
upper panel (dotted curve).  
} 
\end{figure} 
 
Most of the XDINSs have magnetic fields $B \ge 
10^{13}$ G and  color temperatures $\approx 10^6$~K \citep{Haberl:07}. 
Hydrogen model atmospheres are partially ionized under these conditions 
and the vacuum polarization effect is also significant. 
Here we present  first results of modeling of partially ionized 
hydrogen atmospheres using our radiative transfer code. In 
Fig.~\ref{f:fig8} we compare  spectra and temperature 
structures of the partially ionized hydrogen model atmospheres with and 
without the partial mode conversion 
effect. 
When the X-mode (having smaller opacity) partially converts to the 
O-mode in the surface layers of the atmosphere, the energy absorbed by 
the O-mode heats these upper layers. As a 
result, the emergent spectra are closer to the blackbody. 
 
\begin{figure} 
\includegraphics[width=1.0\columnwidth]{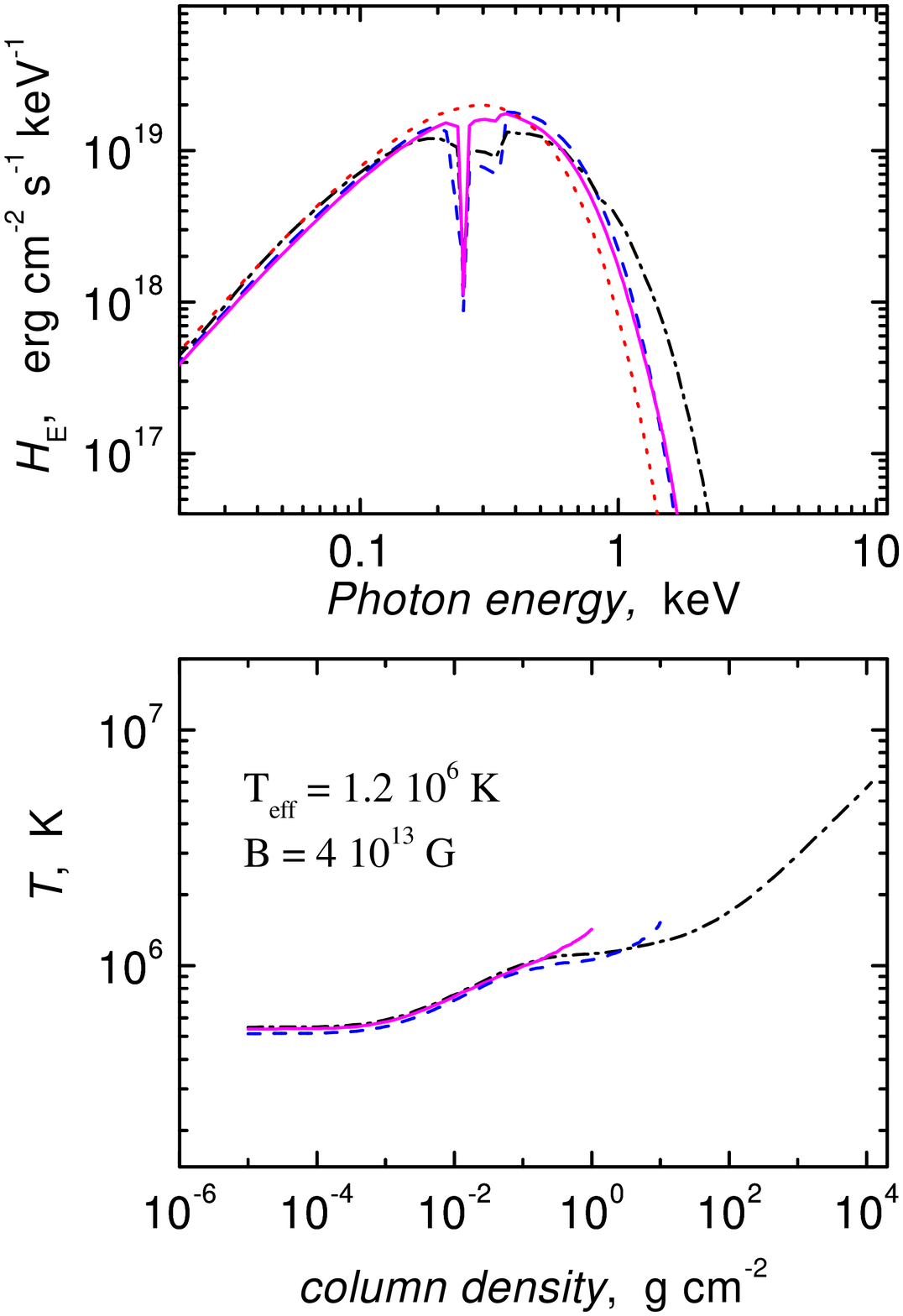} 
\caption{\label{f:fig9} 
Emergent spectra and temperature structures of the partially ionized 
 hydrogen model atmospheres with $T_{\rm eff} = 1.2 \times 
10^6$ K with vacuum polarization 
effect  the partial mode conversion shown for various surface densities $\Sigma$ 
(solid curves - 1 g cm$^{-2}$, dashed curves - 10 g cm$^{-2}$, 
 dash-dotted curves - semi-infinite 
atmosphere).  The magnetic field strength is $B=4 \times 10^{13}$ G.  The corresponding 
blackbody spectrum  is also 
shown in the upper panel (dotted curve). 
} 
\end{figure} 
 
\begin{figure} 
\includegraphics[width=1.0\columnwidth]{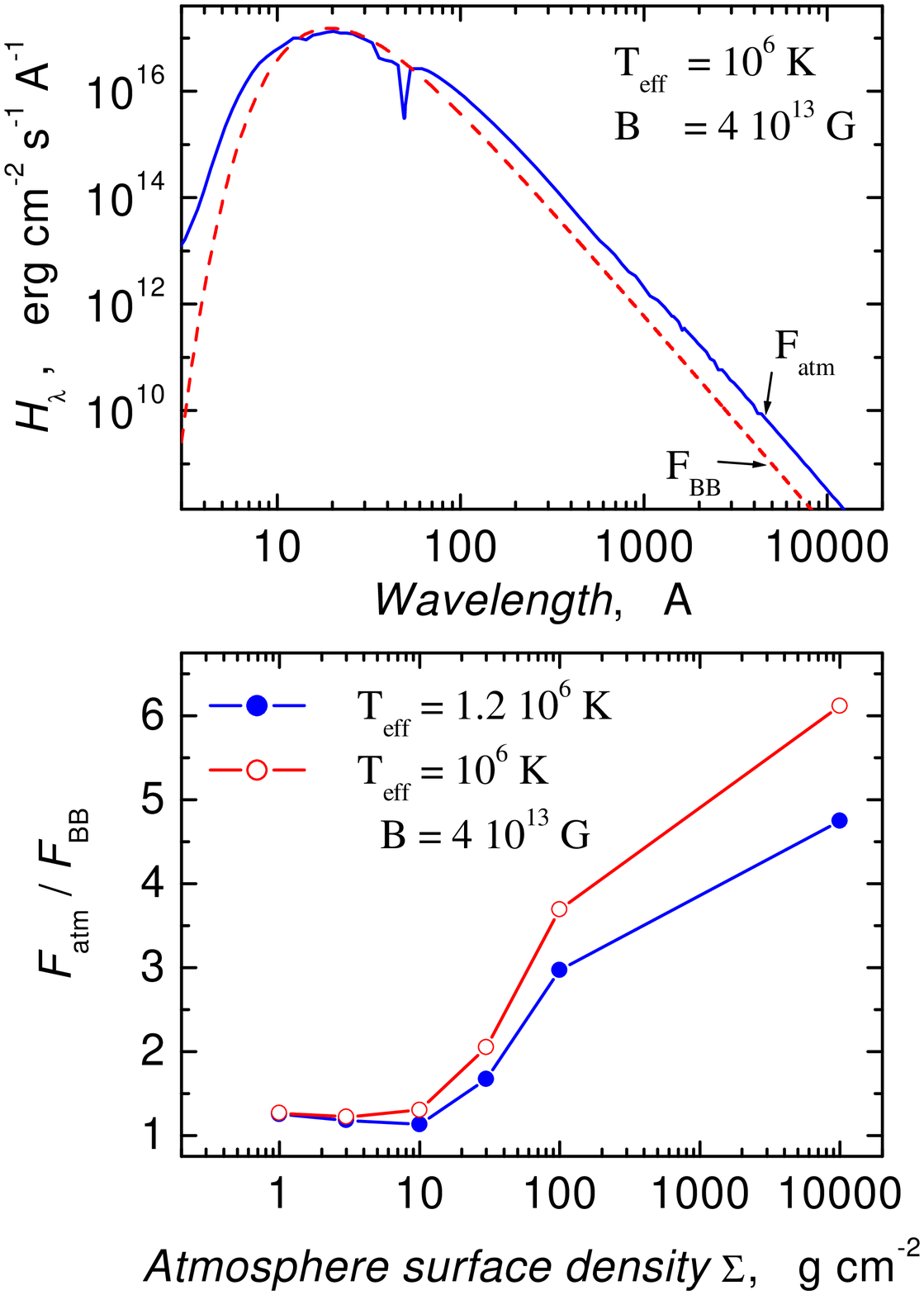} 
\caption{\label{f:fig10} 
{\it Top panel:} Emergent spectrum of the partially ionized hydrogen model 
atmosphere of neutron stars with $T_{\rm eff} = 
10^6$ K with vacuum polarization 
effect,  partial mode conversion and  magnetic field strength $B=4 \times 
10^{13}$ G.  The blackbody spectrum fitted to  the maximum flux of the 
spectral distribution is also shown (dashed curve). At the optical band the model 
atmosphere flux in a few times larger than the blackbody flux. 
{\it Bottom panel:} Ratios of the model atmosphere flux to the blackbody 
(X-ray fitted) flux at the optical band depending on the model atmosphere thickness 
(surface density $\Sigma$) for the models with $T_{\rm eff} = 10^6$ K  and 
$T_{\rm eff} = 1.2 \times 10^6$ K. 
} 
\end{figure}

For some of the XDINSs, optical counterparts have been found 
\citep{Mignanietal:07}. The observed optical fluxes are a 
few times larger compared to the blackbody extrapolation from X-rays 
to the optical range 
(see top panel of Fig.~\ref{f:fig10} for illustration). 
\cite{Hoetal:07} demonstrated, that a single partially ionized thin 
hydrogen atmosphere can explain this problem in the case of brightest 
isolated NS \object{RX\,J1856.4$-$3754}: the model fits well both, 
the observed optical flux and the X-ray spectrum. 
RX\,J1856.4$-$3754 has very low pulsed fraction of radiation 
($\approx 1.2$ \%, 
\citealt{Tie.Meregh:07}), therefore it is possible 
to fit the radiation of this star by the 
single model atmosphere.  Other XDINSs have larger pulse 
fractions, up to 18\% (RBS 1223, \citealt{Haberletal:04}). 
In this case the temperature distribution across the NS surface is not homogenous, 
and the 
excess optical flux can be explained by the radiation from cool surface parts 
\citep{Swopeetal:05}. 
 
We now investigate properties of partially ionized hydrogen modes,
 which can be applied to the \object{RBS 1223} atmosphere. The color 
temperature of this star,  found from  X-ray spectra fitting, 
is close 
to $10^6$ K, with magnetic field $B\approx 4 \times 10^{13}$ G 
\citep{Swopeetal:07}.  In particular, we investigate the optical flux excess 
in comparison 
to the X-ray fitted blackbody flux in this kind of models. 
 For this aim we have calculated two sets of  
 models with vacuum polarization and partial mode 
conversion. The models in the first set have 
effective temperatures $T_{\rm eff} = 10^6$ K and the models of second one 
have effective temperatures $T_{\rm eff} = 1.2 \times 10^6$ K. In both 
sets $B=4 \times 10^{13}$ G, and models with 
surface densities $\Sigma$ = 1, 3, 10, 30, 100 and 10$^5$ (semi-infinite 
model) g cm$^{-2}$ are computed.  In Fig.~\ref{f:fig9} we show  emergent spectra 
and  temperature structures for some models from the second set. 
Clearly, the X-ray spectra of the models with $\Sigma \le 
10$ g cm$^{-2}$ are close to a blackbody and, therefore, better fit 
the observed X-ray spectrum. 
 
 In Fig.~\ref{f:fig10} (bottom panel) we show 
the ratio of the model 
atmosphere flux to the X-ray fitted blackbody flux in the optical band 
depending on $\Sigma$ for 
both sets. 
The observed ratio is about 5 \citep{Kaplanetal:02b}, in
agreement with the semi-infinite atmosphere models. However, the
observed blackbody like X-ray spectrum agrees with the thin
atmosphere models, for which this ratio is close to 1. Therefore,
the observed optical excess cannot be explained by the thin
atmosphere model alone; instead, it can arise due to a nonuniform
surface temperature distribution, in agreement with the RBS 1223
light curve modeling  \citep{Swopeetal:05}.

 
\section{Conclusions} 
\label{s:conclusions} 
 
In this paper a new code for the computation of magnetized 
NS model atmosphere is presented. It can model fully ionized 
 and  partially ionized hydrogen atmospheres in a wide range 
of effective temperatures ($3 \times 10^5$ -- 10$^7$ K) and magnetic 
fields (10$^{12}$ -- 10$^{15}$ G), with any inclination 
of the magnetic 
field to the stellar surface.  The vacuum polarization effect with 
partial mode conversion is taken into consideration. Calculated 
emergent spectra and temperature structures of the model 
atmospheres agree with previously published ones. 
 
We presented new results  obtained using this code. 
We have studied the 
properties of  thin atmospheres above condensed NS surfaces. We 
demonstrated that the proton cyclotron absorption line disappears in the 
thin hydrogen model atmospheres. A new thin ``sandwich'' model atmosphere 
(hydrogen layer above helium layer)  is proposed to explain the occurrence 
of two absorption features in the observed X-ray spectra of some isolated NSs. 

 We analyzed the optical excess (relative to
the X-ray fitted blackbody flux) in the model spectra of
partially ionized hydrogen atmospheres with vacuum polarization
and partial mode conversion. 

A set of model atmospheres  with  parameters  (effective 
temperature and the magnetic field strength) close to the probable parameters 
of the isolated NS RBS\,1223  were calculated. 
We found the optical flux excess $\approx 5$ for the semi-infinite model 
atmospheres decreases down to 1 with decreasing surface density $\Sigma$ 
of the atmosphere. Spectra of thin model atmospheres fit the observed 
RBS\,1223 X-ray spectrum better, therefore we conclude that the observed optical 
excess should be explained by nonuniform surface temperature 
distribution. 
 
The accuracy of the thin and sandwich 
model atmospheres is currently limited  by the inner 
boundary condition for the radiation transfer equation. We used blackbody 
radiation as this condition, but a higher accuracy can be 
achieved 
by replacing it by the condensed surface condition 
\citep{vAetal:05}. 
This will be done in a future work. 
We are also planning to include the effect 
of magnetic field and temperature distributions over the stellar surface
to compute an integral emergent spectrum  from isolated NSs. 
 
\begin{acknowledgements} 
 
VS thanks DFG for financial support (grant We 1312/35-1 and grant 
SFB/Transregio 7 "Gravitational Wave Astronomy") and the President's 
programme for support of leading science schools (grant NSh-4224.2008.2). 
The work of AYP is supported by RFBR grants 05-02-2203 and 08-02-00837 
 and the President's 
programme for support of leading science schools (grant NSh-2600.2008.2). 
 
\end{acknowledgements}


\bibliographystyle{aa} 

\begin{thebibliography}{} 
 
\bibitem[{{Baiotti} {et~al.}(2008) {Baiotti}, {Giacomazzo} \& {Rezzolla}}] 
{Baiottietal:08} 
Baiotti, L., Giacomazzo, B. \& Rezzolla, L. 2008, 
Phys. Rev. D, 78, 084033 
 
\bibitem[{{Brown} {et~al.}(2002){Brown}, {Bildsten}, \& {Chang}}]{BBC:02} 
{Brown}, E.~F., {Bildsten}, L,, \& {Chang}, P. 2002, 
\apj, 574, 920 
 
\bibitem[{{Burwitz} {et~al.}(2001){Burwitz}, {Zavlin}, {Neuh\"auser} {et~al.}}] 
{Burwitzetal:01} {Burwitz}, V., {Zavlin}, V.E., {Neuh\"auser}, R. et~al. 
2001, A\&A, 379, L35 
 
\bibitem[{{Burwitz} {et~al.}(2003){Burwitz}, {Haberl}, {Neuh\"auser} {et~al.}}] 
{Burwitzetal:03} {Burwitz}, V., {Haberl}, F., {Neuh\"auser}, R. et~al. 
2003, A\&A, 399, 1109 
 
 
 
\bibitem[{{G\"ansicke} {et~al.}(2002){G\"ansicke}, {Braje}, \& 
  {Romani}}]{Gansickeetal:02} 
{G\"ansicke}, B.~T., {Braje}, T.~M., \& {Romani}, R.~W. 2002, \aap, 386, 1001 
 
\bibitem[{{Ginzburg}(1970)}]{Ginzburg:70} 
{Ginzburg}, V.L. 1970, {The Propagation of Electromagnetic Waves in Plasmas}, 
(2nd ed. Oxford: Pergamon) 
 
\bibitem[{{Haberl} {et~al.}(2004) {Haberl}, {Motch}, {Zavlin} {et~al.}}] 
{Haberletal:04} 
{Haberl}, F., {Motch}, C., {Zavlin}, V.E. {et~al.} 2004, \aap, 424, 635 
 
\bibitem[{{Haberl}(2007)}]{Haberl:07} 
{Haberl}, F. 2007, A\&SS, 308, 181 
 
\bibitem[{{Ho} \& {Lai}(2001)}]{Ho.Lai:01} 
{Ho}, W.~C.~G. \& {Lai}, D. 2001, \mnras, 327, 1081 
 
\bibitem[{{Ho} \& {Lai}(2003)}]{Ho.Lai:03} 
{Ho}, W.~C.~G. \& {Lai}, D. 2003, \mnras, 338, 233 
 
\bibitem[{{Ho} \& {Lai}(2004)}]{Ho.Lai:04} 
{Ho}, W.~C.~G. \& {Lai}, D. 2004, \apj, 607, 420 
 
\bibitem[{{Ho} {et~al.}(2007){Ho}, {Kaplan}, {Chang}, {van Adelsberg} 
\& {Potekhin}}]{Hoetal:07} 
{Ho}, W.~C.~G.,  {Kaplan}, D.L., {Chang}, P., {van Adelsberg}, M. \& {Potekhin}, 
 A.Y. 2007, \mnras, 375, 821 
 
\bibitem[{{Ho} {et~al.}(2008){Ho}, {Potekhin} \& {Chabrier}}]{Hoetal:08} 
{Ho}, W.~C.~G., {Potekhin}, A.Y. \& {Chabrier}, G. 
2008, \apjs, 178, 102 
 
\bibitem[{{Ibragimov} {et~al.}(2003){Ibragimov}, {Suleimanov}, {Vikhlinin}, \& 
  {Sakhibullin}}]{Ibragimov.etal:03} 
{Ibragimov}, A.~A., {Suleimanov}, V.~F., {Vikhlinin}, A., \& {Sakhibullin}, 
  N.~A. 2003, Astronomy Reports, 47, 186 
 
\bibitem[{{Kaminker} {et~al.} (1982){Kaminker}, {Pavlov}, \& {Shibanov}}] 
{Kaminkeretal:82} {Kaminker}, A.D.,  {Pavlov}, G.G. \& {Shibanov} Yu.A. 
 1982, Ap\&SS, 86, 249 
 
\bibitem[{{Kaminker} {et~al.} (1983){Kaminker}, {Pavlov}, \& {Shibanov}}] 
{Kaminkeretal:83} {Kaminker}, A.D.,  {Pavlov}, G.G. \& {Shibanov} Yu.A. 
 1983, Ap\&SS, 91, 167 
 
\bibitem[{{Kaplan} {et~al.} (2002a){Kaplan}, {van Kerkwijk}, {Anderson}}] 
{Kaplanetal:02a} {Kaplan}, D.L., {van Kerkwijk}, M.H. {Anderson}, J. 
 2002a, \apj, 571, 447 
 
\bibitem[{{Kaplan} {et~al.} (2002b){Kaplan}, {Kulkarni} \& {van Kerkwijk}}] 
{Kaplanetal:02b} {Kaplan}, D.L.,  {Kulkarni}, S.R. \& {van Kerkwijk}, M.H. 
 2002b, \apj, 579, L29 
 
\bibitem[{{Kaplan} {et~al.} (2003){Kaplan}, {van Kerkwijk}, {Marshall} {et~al.}}] 
{Kaplanetal:03} {Kaplan}, D.L., {van Kerkwijk}, M.H., {Marshall}, H.L., et al. 
 2003, \apj, 590, 1008 
 
\bibitem[{{Kaspi}(2007)}]{Kaspi:07} 
{Kaspi}, V.M. 2007, A\&SS, 308, 1 
 
\bibitem[{{Kurucz}(1970)}]{Kurucz:70} 
{Kurucz}, R.~L. 1970, SAO Special Report, 309 
 
\bibitem[{{Kurucz}(1993)}]{Kurucz:93} 
{Kurucz}, R. 1993, Atomic data for opacity calculations.~Kurucz CD-ROMs, 
  Cambridge, Mass.: Smithsonian Astrophysical Observatory, 1993, 1 
 
\bibitem[{{Lai}(2001)}]{Lai:01} 
{Lai}, D. 2001, Reviews of Modern Physics, 73, 629 
 
\bibitem[{{Lai} \& {Salpeter}(1997)}]{Lai.Salpeter:97} 
{Lai}, D. \& {Salpeter}, E.~E. 1997, \apj, 491, 270 
 
\bibitem[{{Lai} \& {Ho}(2002)}]{Lai.Ho:02} 
{Lai}, D. \& Ho. W.C.G. 
2002, \apj, 566, 373 
 
\bibitem[{{Lai} \& {Ho}(2003)}]{Lai.Ho:03} 
{Lai}, D. \& Ho. W.C.G. 
2003, \apj, 588, 962 
 
\bibitem[{{Lattimer} \& {Prakash}(2007)}]{LP07} 
{Lattimer}, J.M. \& {Prakash}, M. 2007, 
Phys.\ Rep., 442, 109 
 
\bibitem[{{Medin} \& {Lai}(2007)}]{M07} 
{Medin}, Z. \& {Lai}, D. 2007, 
\mnras, 382, 1833 
 
\bibitem[{{Mereghetti}(2008)}]{Mereghetti:08} 
{Mereghetti}, S. 2008, 
\aapr, 15, 225 
 
\bibitem[{{Mereghetti} {et~al.}(2002){Mereghetti}, {Tiengo}, \& 
  {Israel}}]{Mereghettietal:02} 
{Mereghetti}, S., {Tiengo}, A., \& {Israel}, G.~L. 2002, \apj, 569, 275 
 
\bibitem[{{Mereghetti} {et~al.}(2007){Mereghetti}, {Esposito} 
\& {Tiengo}}]{Mereghettietal:07} 
{Mereghetti}, S., {Esposito}, P. \& {Tiengo}, A. 2007, 
 A\&SS, 308, 13 
 
\bibitem[{{M\'es\'zaros}(1992)}]{Mezharos:92} 
{M\'es\'zaros}, P. 1992, {High-Energy Radiation from Magnetized Neutron Stars} 
(Chicago, Univ. Chicago Press) 
 
\bibitem[{{Mignani} {et~al.}(2007){Mignani}, {Bagnulo}, 
{De Luca}{et~al.}}]{Mignanietal:07} 
{Mignani}, R.P,  {Bagnulo}, S., {De Luca}, A. et~al. 
 2007,  A\&SS, 308, 203 
 
\bibitem[{{Mihalas}(1978)}]{Mihalas:78} 
{Mihalas}, D. 1978, {Stellar atmospheres, 2nd edition} (San Francisco, 
  W.~H.~Freeman and Co.) 
 
\bibitem[{{Mori} \& {Ho}(2007)}] 
{MH:07} {Mori}, K. \& {Ho}, W.C.G. 
2007, \mnras, 377, 905 
 
\bibitem[{{Motch} {et~al.}(2003) {Motch}, {Zavlin} \& {Haberl}}] 
{Motchetal:03} {Motch}, C., {Zavlin}, V.E. \& {Haberl}, F. 
2003, A\&A, 408, 323 
 
\bibitem[{{Olson} \& {Kunasz}(1987)}] 
{Ols.Kun:87} {Olson}, G.L. \& {Kunasz}, P.B. 
1987, JQSRT, 38, 325 
 
\bibitem[{{\"Ozel}(2001)}]{Ozel:01} 
{\"Ozel}, F. 2001, \apj, 563, 276 
 
 
\bibitem[{{Pavlov} \& {Gnedin}(1984)}]{PG:84} 
{Pavlov}, G.~G., \& {Gnedin}, Yu.~N. 1984, 
Astrophys.\ Space Phys.\ Rev., 3, 197 
 
\bibitem[{{Pavlov} \& {M\'esz\'aros}(1993)}]{PM:93} 
{Pavlov}, G.~G., \& {M\'esz\'aros}, P. 1993, 
\apj, 416, 752 
 
\bibitem[{{Pavlov} {et~al.}(1994){Pavlov}, {Shibanov}, {Ventura} \& {Zavlin}}] 
 {Pavlovetal:94} {Pavlov}, G.G., {Shibanov}, Yu.A., {Ventura}, J. 
\& {Zavlin}, V.E. 1994, \aap, 289, 837 
 
\bibitem[{{Pavlov} {et~al.}(2002){Pavlov}, 
  {Sanwal}, {Garmire}, \& {et al.}}]{Pavlovetal:02a} 
{Pavlov}, G.~G., {Sanwal}, D., {Garmire}, G.P.,  2002, In: Slane, P.O., 
Gaensler, B.M. (eds) Neutron Stars in Supernova Remnants. ASP Conf. Ser. 271, 
247 
 
\bibitem[{{Pavlov} {et~al.}(2004){Pavlov}, 
  {Sanwal}, {Garmire}, \& {et al.}}]{Pavlovetal:04} 
{Pavlov}, G.~G., {Sanwal}, D., {Teter}, M.A.,  2004, in 
Young Neutron Stars and Their Enviroments (Proceedings of the 
IAU Symp. 218), ed.\ F.~Camilo \& B.~M. Gaensler 
(ASP, San Francisco), 239 
 
\bibitem[{{P\'erez-Azor\'{\i}n} {et~al.}(2005){P\'erez-Azor\'{\i}n}, {Miralles}, \& {Pons}}]{PAMP05} 
{P\'erez-Azor\'{\i}n}, J.~F., {Miralles}, J.~A., \& {Pons}, J.~A. 
2005, 
\aap, 433, 275 
 
\bibitem[{{Pons} {et~al.}(2002){Pons}, {Walter}, {Lattimer}, {Prakash}, 
  {Neuh\"auser}, \& {An}}]{Ponsetal:02} 
{Pons}, J.~A., {Walter}, F.~M., {Lattimer}, J., {et~al.} 2002, 
\apj, 564, 981 
 
\bibitem[{{Potekhin} \& {Chabrier}(2003)}]{PCh:03} 
{Potekhin}, A.Y. \& {Chabrier} G. 
2003, \apj, 585, 955 
 
\bibitem[{{Potekhin} \& {Chabrier}(2004)}]{PCh:04} 
{Potekhin}, A.Y. \& {Chabrier} G. 
2004, \apj, 600, 317 
 
\bibitem[{{Potekhin} {et al.}(1999){Potekhin}, {Chabrier} \& 
{Shibanov}}]{PCS:99} 
{Potekhin}, A.Y., {Chabrier} G.\& {Shibanov}, Yu.~A. 
1999, \pre, 60, 2193 
 
\bibitem[{{Potekhin} {et al.}(2004){Potekhin}, {Lai}, {Chabrier} \& {Ho}}] 
{Potekhinetal:04} 
{Potekhin}, A.Y., {Lai}, D., {Chabrier} G., \& {Ho}, W.~C.~G. 
2004, \apj, 612, 1034 
 
\bibitem[{{Rajagopal} \& {Romani}(1996)}]{Rajagopal.Romani:96} 
{Rajagopal}, M. \& {Romani}, R.~W. 1996, \apj, 461, 327 
 
\bibitem[{{Rajagopal} {et~al.}(1997){Rajagopal}, {Romani}, \& 
  {Miller}}]{Rajagopaletal:97} 
{Rajagopal}, M., {Romani}, R.~W., \& {Miller}, M.~C. 1997, \apj, 479, 347 
 
\bibitem[{{Rauch} {et~al.}(2008){{Rauch}, Suleimanov} \& {Werner}}] 
{Rauchetal:08} 
{Rauch}, T., {Suleimanov}, V. \& {Werner}, K. 
  2008, \aap, 490, 1127 
 
\bibitem[{{Romani}(1987)}]{Romani:87} 
{Romani}, R.~W. 1987, \apj, 313, 718 
 
\bibitem[{{Sanwal} {et~al.}(2002){Sanwal}, {Pavlov}, {Zavlin}, \& 
  {et~al.}}]{Sanwaletal:02} 
{Sanwal}, D., {Pavlov}, G.G., {Zavlin}, V.E. et. al. 
2002, \apj, 574, L61 
 
\bibitem[{{Schwope} {et~al.}(2005){Schwope}, {Hambaryan}, {Haberl} 
 {et~al.}}]{Swopeetal:05} 
{Schwope}, A.D, {Hambaryan}, V., {Haberl}, F.  {et~al.} 
2005,  \aap, 441, 597 
 
\bibitem[{{Schwope} {et~al.}(2007){Schwope}, {Hambaryan}, {Haberl} 
\& {Motch}}]{Swopeetal:07} 
{Schwope}, A.D, {Hambaryan}, V., {Haberl}, F. \& {Motch}, C. 
2007,  A\&SS, 308, 619 
 
\bibitem[{{Shibanov} {et~al.}(1992){Shibanov}, {Zavlin}, {Pavlov}, \& 
  {Ventura}}]{Shibanovetal:92} 
{Shibanov}, I.~A., {Zavlin}, V.~E., {Pavlov}, G.~G., \& {Ventura}, J. 1992, 
  \aap, 266, 313 
 
\bibitem[{{Suleimanov} \& {Poutanen}(2006)}]{Sul.Pout:06} 
{Suleimanov}, V. \& {Poutanen}, J. 2006, \mnras, 369, 2036 
 
\bibitem[{{Suleimanov} \& {Werner}(2007)}]{Sul.Wer:07} 
{Suleimanov}, V. \& {Werner}, K. 
  2007, \aap, 466, 661 
 
\bibitem[{{Swartz} {et~al.}(2002){Swartz}, {Ghosh}, {Suleimanov}, {Tennant}, \& 
  {Wu}}]{Swartz.etal:02} 
{Swartz}, D.~A., {Ghosh}, K.~K., {Suleimanov}, V., {Tennant}, A.~F., \& {Wu}, 
  K. 2002, \apj, 574, 382 
 
\bibitem[{{Tiengo} \& {Mereghetti}(2007)}]{Tie.Meregh:07} 
{Tiengo}, A. \& {Mereghetti}, S. 2007, \apj, 657, L101 
 
\bibitem[{{Tr\"umper} {et~al.}(2004) {Tr\"umper}, {Burwitz}, {Haberl} \& {Zavlin}}]{Trumperetal:04} 
 {Tr\"umper}, J.E., {Burwitz}, V., {Haberl}, F. \& {Zavlin}, V.E.  2004, 
Nuclear Phys. B Proc. Suppl., 132, 560 
 
\bibitem[{{Turolla} {et~al.}(2004){Turolla}, {Zane}, \& 
{Drake}}]{TZD:04} 
{Turolla}, R., {Zane}, S., \& 
{Drake}, J.~J. 2004, 
\apj, 603, 265 
 
\bibitem[{{van Adelsberg} \& {Lai}(2006)}]{vAL:06} 
{van Adelsberg}, M. \& {Lai}, D. 
 2006,  \mnras, 373, 1495 
 
\bibitem[{{van Adelsberg} {et al.}(2005){van Adelsberg},  
{Lai}, {Potekhin} \& {Arras}}]{vAetal:05} 
{van Adelsberg}, M., {Lai}, D., {Potekhin}, A.Y. \& {Arras}, P. 
 2005,  \apj, 628, 902 
 
\bibitem[{{van Kerkwijk} \& {Kaplan}(2007)}]{vKK:07} 
{van Kerkwijk}, M.~H., \& {Kaplan}, D.~L. 
 2007,  A\&SS, 308, 191 
 
\bibitem[{{Ventura}(1979)}]{Ventura:79} 
{Ventura}, J. 1979, 
\prd, 19, 1684 
 
\bibitem[{{Werner} \& {Deetjen}(2000)}]{Werner:00} 
{Werner}, K., \& {Deetjen}, J. 2000, in Pulsar Astronomy -- 2000 and Beyond, 
ed.\ M.~Kramer, N.~Wex, \& R.~Wielebinski, 
ASP Conf.\ Ser., 202, 623 
 
\bibitem[{{Zane} {et~al.}(2001)}]{Zaneetal:01} 
Zane, S., Turolla, R., Stella, L., \& Treves, A. 2001, 
\apj, 560, 384 
 
\bibitem[{{Zavlin}(2009)}]{Zavlin:09} 
{Zavlin}, V.~E. 2009, Theory of radiative transfer 
in neutron star atmospheres and its applications, 
in Neutron Stars and Pulsars (Proceedings of the 
  363.\ WE-Heraeus Seminar),   
  ed.\ W.~Becker (Springer, New York), 181 
 
\bibitem[{{Zavlin} {et~al.}(1996){Zavlin}, {Pavlov}, \& 
  {Shibanov}}]{Zavlinetal:96} 
{Zavlin}, V.~E., {Pavlov}, G.~G., \& {Shibanov}, I.~A. 1996, \aap, 315, 141 
 
 
\end{thebibliography}

\end{document}